# Dynamic modelling of a freshwater stream community from experimental field data

Roberto de Moraes Lima Silveira[1], Claudio Gonçalves Carvalhaes[2], Claudio Sasaki[3], Felipe Augusto Maurin Krsulovic[4] & Timothy Peter Moulton[5]

[1]Universidade Federal de Mato Grosso, Institute of Biosciences, Av. Fernando Correia da Costa, Cuiabá, 78060-900, CCBS III sala 3, Mato Grosso, Brazil.
[2] Center for the Study of Language and Information, Stanford University, 210 Panama St, Stanford, CA 94305-4101.
[3]Universidade do Estado do Rio de Janeiro, and Institute of Mathematics and Statistics, Rua São Francisco Xavier, 524, Rio de Janeiro, Brazil.
[4]Facultad de Ciencias Biológicas, Pontifícia Universidad Católica de Chile, Alameda 340, Santiago, Chile.
[5]Universidade do Estado do Rio de Janeiro, and Institute of Biology, Rua São Francisco Xavier, 524, Rio de Janeiro, Brazil.
E-mails: silveira@ufmt.br, c.g.carvalhaes@gmail.com, famk@globo.com, moulton@uerj.br

We developed mathematical models that simulate community dynamics derived from a series of perturbation experiments. These experiments were performed in an Atlantic Forest stream. The three trophic level community was submitted to two combinations of press perturbation experiment. In the first, the top trophic level was removed by electrical exclusion. In the second configuration, the top trophic level was still excluded plus a group of species from the second trophic level was inhibited, also by electrical pulses. Experiments were repeated at different times to increase confidence in the observed mechanisms. Community responses indicated the existence of cascading interactions similar to a classic trophic cascade. The community was composed of *Macrobrachium* shrimps at the top trophic level, Ephemeroptera and Chironomidae larval insects at the second trophic level and periphyton (= epilithon = biofilm) at the first trophic level. The shrimp exclusion caused an increase in ephemeropterans and chironomids and led the periphyton to decrease. When ephemeropterans were inhibited together with the exclusion of shrimps, the chironomids and the periphyton increased. Although the insects-periphyton interactions were of a trophic nature, the shrimps-insects interactions were not. We modelled the community interactions by means of differential equation systems, simulating the three configurations: 1. natural condition, 2. shrimp exclusion condition and 3. shrimp-exclusion-plus-ephemeropteran-inhibition condition. All parameters were calculated via experimental results and the stability of the models was determined by their matrix eigenvalues. The models successfully reproduced the qualitative responses of the community. They proved to have attraction points, which enables us to predict that the system has some ability to return to its previous condition after local perturbation.

Keywords: aquatic; trophic cascade; mathematical model; electrical exclusion; periphyton

## INTRODUCTION

Ecological studies are generally meant to characterize interactions which later allow us to develop an idea of a whole process or system. However exclusively intuitive models are often not enough for this understanding (Hannon & Ruth 2001). Some of the difficulties lie in the nature of the non-linear behaviour which is frequently found in population dynamics, interaction coefficients and higher order interactions (Abrams 1993, Closs *et al*. 1993 and Billick & Case 1994).

In an attempt to overcome the limitations of intuitive modelling, biological processes have been described mathematically for more than one century (Muller & Joshi 2000). A mathematical model does not offer just a more consistent idea but also can simulate predictions based on our premises. This allows us to re-evaluate our understanding and rebuild a more coherent model (Hannon & Ruth 2001).

Traditionally mathematical modelling is well accepted in population dynamics (Volterra 1937, Murdoch 1994, Dennis *et al*. 2001), more recently in food web studies (Pimm 1982, Closs *et al*. 1993 Carpenter *et al*. 1994, McCann & Yodzis 1994, Berryman *et al*. 1995, Vander Zanden & Rasmussen 1996) and community dynamics (May 1973, Pimm 1991, Berryman *et al*. 1995, Laska & Wootton 1998, Silveira & Moulton 2000). Focusing on the last two areas, Vander Zanden & Rasmussen (1996) state the three major objectives of trophic level interaction modelling: 1. to search consistent patterns of community structure, 2. to study factors structuring communities and 3. to study energy flow.

We will concentrate on the factors that structure communities; generally these studies are carried out by experimental manipulations. Although experimental approaches have been criticized for their inability to create a reliable data set (Yodzis 1988), this method allows us to investigate indirect effects that otherwise could be easily overlooked. Experimental manipulations also provide good material for modelling. First, experiments can show the community dynamics which can be incorporated into a preliminary model and a set of parameters. Then experiments can be rearranged to obtain the model parameters.

### *COMMUNITY INTERACTIONS AND EVIDENCE OF TROPHIC CASCADES*

Direct trophic interactions are not the sole processes that might structure a community; indirect interactions are often important system components as well. Wootton (1994) describes a series of indirect effects in ecological communities such as apparent competition, indirect mutualism and higher order interactions and shows how to detect them. He states that these effects have been commonly detected by perturbation experiments but that most authors may not be aware that indirect effects could be overlooked when they occur at the same time and in the same direction as a direct effect. Usually in these circumstances direct effects are imputed.

Indirect effects such as trophic cascades (Strong 1992, Pace *et al*. 1999) are especially present in aquatic ecosystems. A trophic cascade occurs when a top predator has an indirect interference in one or more species at the bottom of the food chain. Holt (2000) comments that terrestrial ecosystems could be less exposed to trophic cascades because only a few plant species would be seriously affected by a lack of predators. Generally trophic cascades are expected to occur in conditions where primary producers are easily submitted to herbivory pressure, only a few species act as grazers and a few others act as predators. And finally trophic cascades generally occur in places that are isolated in space (Strong 1992).

Various trophic cascades in temperate and tropical streams have been described (Strong 1992, Pace *et al*. 1999). Power (1984 and 1990) and Power *et al*. (1985) found trophic cascades involving sediments, fishes and birds and others involving algae, aquatic insects and fishes. Lodge *et al*. (1994) also found a trophic cascade involving algae, snails and crayfishes. Kneib (1988) describes a trophic cascade involving aquatic insects, freshwater shrimps and fishes.

### *THE STREAM COMMUNITY AT ILHA GRANDE*

The situation at our study site was similar to that suggested by Strong (1992) as where a trophic cascade could be expected. Periphyton was potentially highly productive and easily exposed to gazers; there were only a few grazers and only one predator. So a trophic cascade was expected and it could be modelled mathematically through our perturbation experiments. Evidences of trophic cascades were verified in two Ilha Grande Streams (Moulton *et al*. 2010). On the other hand, tropical stream communities composed of shrimps and macroinvertebrates did not show tropic cascades in experiments in Puerto Rico (Pringle and Blake 1994) and Costa Rica (Pringle & Hamazaki 1998).

Our objectives are to define the main community interactions and to develop a useful model of community dynamics using results from perturbation experiments.

The three trophic level community was composed of shrimps at the top level, aquatic insects at the second trophic level and periphyton at the first trophic level. Based on the community response to the manipulations Silveira & Moulton (2000) were able to describe the community dynamics and produce a simple model of this system. Although the community dynamics was well described qualitatively, the mathematical model was not adequate to explain the processes observed during the experiments. This model was relevant to long term community dynamics, involving reproductive events, but did not simulate the interactions of short term experiments.

We remodelled the community dynamics using a different set of differential equations to better describe the community under experimental conditions. This new model aimed to incorporate the interactions revealed by experiments and describe the community dynamics based on them. We used the results of manipulation experiments to calculate the growth rates and interaction coefficients needed for modelling. Although modelling ecological interactions, populations and community dynamics are not unusual (May 1972, Pimm 1991, Hastings *et al*. 1993, Abrams 1993, Closs *et al* 1993, Conroy *et al*. 1995, Berrryman *et al*. 1995, Leibold 1996) only a few studies employ parameters estimated from experimental or observational data set (e.g. Wootton 1994, Schmitz 1997, Wootton 1997, Laska and Wootton 1998 and Dennis *et al* 2001).

The models simulated organism density in an area with the same dimensions as that used in experiments. Population density of the different organisms changes when community composition is changed. We simulated the three community compositions observed, first natural conditions, then with shrimps excluded and finally with shrimps excluded and ephemeropterans inhibited. We also tested stability in the vicinity of the system's equilibrium point by analysing the eigenvalues of the community matrix.

## METHODS

### *STUDY SITE*

Experiments were performed at Córrego da Andorinha stream which is located near Vila Dois Rios village, Ilha Grande island Rio de Janeiro State, Brazil. Córrego da Andorinha is a low order stream on the oceanic side of the island; we carried out our experiments at a particular site which was ~6 km from the source and ~1 km from the mouth (23° 10.97' S, 44° 12.08' W, 70 m asl). Discharge was approximately 200 L/s at base flow. The experimental area was ~28 $m^2$ of horizontal bed rock, and depth varied between 26.6 cm to 3.5 cm. Immediately upstream there is a pool (20 m diameter and 3.5 deep) and downstream the bedrock inclines ~30° forming a waterfall (Silveira & Moulton 2000; Moulton *et al*. 2004).

### *THE COMMUNITY*

The community in the experimental area was relatively simple with only a few taxa for each main group. The two species of shrimps were *Macrobrachium olfersi* (Wiegmann 1836) and *Potimirim glabra* (Müller 1857). However, during the experiments predominantly *M. olfersi* was observed in the area. Ephemeropterans (Ephemeroptera, Baetidae) were represented by three morphospecies, *Cloeodes* sp., *Americabaetis* sp. and *Baetodes* sp. Chironomids (Order Diptera) belonged to subfamily Tanypodinae (tribe Pentaneurini- one morphospecies), subfamily Orthocladiinae (*Cricotopus* sp- one morphospecies), subfamily Chironominae, tribe Tanytarsini- one morphospecies and tribe Chironomini, *Beardius* sp. also one morphospecies. The algae assemblage was principally composed of diatoms *Cymbella* sp. and *Fragilaria* sp; blue green algae *Rivularia* sp and *Oscillatoria* sp. and green algae *Zygnema* sp. and *Cladophora* sp.

This community was distributed on a bedrock substrate without cobbles, ranging from 5 cm to 30 cm depth and well exposed to sunlight. The exposure and lack of refuges could explain why the only fish observed in the vicinity (Crenuchidae, *Characidium japuhybensis* Travassos 1949) was absent in that particular place and why other groups of aquatic insects such as Trichoptera and Plecoptera were also absent.

*EXPERIMENTS IN THE FIELD*

The community was submitted to five press perturbation experiments where faunal components were excluded and or inhibited for a period of time. Field exclusion experiments in streams traditionally use cages to create the exclusion treatment (Diehl 1995 and Flecker 1996). We chose to create exclusion treatments using electricity to reduce artificial effects caused by cages following the principles developed by Pringle and Blake (1994); further details are found in Silveira & Moulton (2000); Moulton *et al*. (2004); Souza & Moulton (2005). Electric pulses were generated from electric fence chargers and applied in the water over given areas by parallel uncovered wires. The arrangement of wires created an electrified corridor that we called exclusion zone. We used two types of electrifiers with different electrical intensity. The Ballerup® (Alfa S.A., São Paulo, Brazil) had a low intensity electrical pulse and the Speedrite® (model SB5000, Tru-Test Ltd., New Zealand) electrifier had a high intensity electrical pulse. Energy to the electric fence was supplied by a 12 V battery, which was recharged by a solar panel.

Electric fields affect animals in proportion to their body sizes; large animals are affected at lower intensities than small animals. We adjusted the electrifiers to perform two types of exclusion: Type I excluded only shrimps and type II with a higher electric intensity excluded shrimps and inhibited ephemeropterans at the same time, without affecting chironomids and periphyton (Fig. 1). As a result of our exclusion method we could compartmentalize the community into four classes with different reactions to the two levels of electrification: shrimps, ephemeropterans, chironomids and periphyton

Each experiment had at least one electrified exclusion zone and a non electrified zone monitored as control. Both electrified and non electrified zones were not replicated within each experiment and the two kinds of manipulation were repeated in time. The lack of replicas does not compromise the major point of this paper which is to study community interactions through provoked responses from perturbation experiments and modelling them mathematically.

Experiments were carried out according to the availability of the field station and of the team workers. This caused some differences mainly in experiment time duration (Table 1). Experiments 4 and 5 were conducted simultaneously (Moulton *et al*. 2004).

Aquatic insects were sampled using a 200 μm mesh Surber sampler on four occasions for the first three experiments and on three occasions for the last experiment. This was done always by the same team member in order to minimize operator error (Table 1). Sampled insects were preserved and counted and taxonomically identified under stereo microscope (40 X). Shrimps were visually counted inside treatment areas to estimate their activities. Unlike Surber sampling, visual counting was not substrate destructive and it could be done every day and night during the experiment. The counting was not done by a fixed person because it was too laborious, but care was taken to minimize operator error.

Periphyton was sampled using a sampler modified from Loeb (1981) and the sample was subsequently divided in three: one part to quantify periphyton total dry mass, another to quantify periphyton chlorophyll *a* and the last part to identify periphyton algae taxonomically. Periphyton dry mass was estimated using a calculated relation between the sample turbidity and its dry mass. Periphyton algae was filtered onto a glass fibre GF/F filter and its Chlorophyll *a* estimated using a spectrophotometer.

We used Analysis of Variance to identify significant differences in the densities of organisms and total dry mass between electrified and non electrified zones. Each group of organisms was analysed separately and the densities between treatments were compared for the last sampling day. Animal abundance data were transformed to Log (x+1), total dry mass and chlorophyll *a* data were transformed to Log (x) before statistical analysis to homogenize variances.

*BUILDING THE MODEL*

We built a system of differential equations that represented community population fluctuations for an area of 1 $m^2$. Unlike traditional Lotka-Volterra models, population increase of animals was not by reproduction since experiments were too short in time. Thus growth rate was not proportional to population density. In our model, the population dynamics came from migration behaviour. The algal population behaved as a Lotka-Volterra-type model (Berryman *et al*. 1995). The total dry mass fluctuation represented its accumulation and loss to detritivores. Unlike the algal population model the total dry mass model considered that detritus had a constant accumulation rate which was not multiplied by any equation variable. Such a system of mixing biomass and individual numbers is known as the "well mixed producer" model (Nisbet *et al*. 1997).

The system was built to have the least number of parameters possible in order to reduce its uncertainty and to model the most important elements of the community. Thus the system has five compartments composed of algae, total dry mass (of periphyton), Chironomids, Ephemeropterans and shrimps. Three different systems were built to represent the system under natural condition, Type I exclusion experiments and Type II exclusion experiments, Fig. 1.

## *ESTIMATING PARAMETERS*

### *Periphyton algae rate of increase ($r_a$) and algae logistic term ($K_a$)*

These parameters were calculated in Type II experiment conditions, where algae grew unconstrained by herbivores. In these conditions algae dynamics was presumed to be ruled by the classic Lotka-Volterra equation. Using the integral form of this equation we calculated the parameters needed:

$$A_{(t)} = \frac{K_a}{1 + e^{b - r_a t}} \qquad \text{Eq. (1)}$$

The parameters were estimated by non-linear regression (NONLIN module of SYSTAT) along with a self consistent test. This test demands a base of previous parameter estimation. $K_a$ was estimated as the natural final algae density shown during experiments. The *b* parameter is:

$$b - r_a t = Ln\left(\frac{K_a - A_{(t)}}{A_{(t)}}\right) \qquad \text{Eq. (2)}$$

Plotting the right side of equation (2) against time the left side is a linear equation. The *b* parameter value is found when time is equal to zero. The $r_a$ parameter was previously estimated from the exponential part of the curve. The NONLIN module used the previously estimated parameters as a base to find its own estimated values. If the NONLIN calculated values were close to those estimated from the exponential curve we had enough confidence to use the values calculated from the curve.

### *Total dry mass increase rate ($r_m$) and logistic term ($K_m$)*

As with the algae, the total dry mass increase rate and its logistic term were calculated from Type II experiment conditions. Unlike the algae equation, the total dry mass dynamics was described by an equation but the increase of this component is not proportional to itself like algae. This represents a constant increase rate independent of the amount of total dry mass:

$$M_{(t)} = K_m\left(1 - e^{-\frac{r_m}{K_m}t}\right) \qquad \text{Eq. (3)}$$

Again the NONLIN module was used to perform a self confidence test with previously estimated parameters and calculate the most appropriate parameters.

### *Chironomid and ephemeropteran increase rate ($r_d$ and $r_e$)*

The rate of increase of both these components was proportional to the density of detritus. This is so because of the kind of increase observed during experiments. Both populations are composed of sexually non-mature individuals thus their increase in density could be only due to migratory movements. The experimental area did not provide any shelter so we consider them to be attracted by the food supply. Both insects are dependent on algae as a food resource so their dynamics were defined as increasing in function of algal availability (Brito *et al*. 2006). Such parameters were calculated from Type II experiment conditions for chironomids and Type I and II experiment conditions for ephemeropterans. Under such circumstances, chironomid and ephemeropteran densities are determined by equations (4) and (5) respectively.

$$\frac{dD_{(t)}}{dt} = r_d A_{(t)} \qquad \text{Eq. (4)}$$

$$\frac{dE_{(t)}}{dt} = r_e A_{(t)} \qquad \text{Eq. (5)}$$

We used the data collected in each sampling day to plot by difference approximation the left side of equations (4) and (5) against the algae densities (*A*) recorded in those days. Using a linear regression we were able to estimate the $r_d$ and $r_e$ parameters.

*Shrimp increase rate ($r_s$)*

To estimate this parameter we used data from in situ counting of shrimps during the day and at night. In such conditions shrimp dynamics is defined by equation (6). This parameter was estimated by the same method we used to estimate chironomid and ephemeropteran increase rate.

The only difference is that shrimps are supposed to be attracted by ephemeropteran densities. Chironomids possibly influenced the dynamics, but our data were not sufficient to support this. Ephemeropterans and shrimps have alternate activity (shrimps appear at night and ephemeropterans by day) which suggests a strong predator-prey relationship between them.

Observed shrimp densities at night were plotted against observed ephemeropteran densities of the previous afternoon. To allow such an approach, only data with a low number of shrimps were used. This situation characterizes the population of shrimps in its lowest levels thus allowing us to consider that any intra-specific relation was of negligible effect. In this case the second term of the shrimp equation was considered zero and we used a linear regression to estimate the parameter $r_s$.

$$\frac{dS_{(t)}}{dt} = r_s E_{(t)} \qquad \text{Eq. (6)}$$

*Interaction coefficients between components*

These parameters were calculated via system equilibrium points. We estimated such coefficients according to the situation. The natural condition was considered to be in its stable equilibrium point just before experiments. Using all experimental data for day zero (before electrification) we calculated the mean equilibrium density for all community components. In such circumstances the component variation rate is zero so we substituted all equilibrium point densities and parameters already estimated in the equations and made the equation equal to zero.

The same procedure was performed to calculate the values of the interaction coefficients for Type I and II experiments where some terms of chironomid and ephemeropteran equations were changed and some terms of algae and total dry mass equations were deleted. In these cases the equilibrium point was determined at the end of the experiments when the community was tending to stop its fluctuation.

*Stability analysis*

Characteristic matrix eigenvalues for the three simulated situations were calculated by MATLAB (MATLAB 1991) in order to test the model's stability and then the models were run in Model Maker computer program (SB Technology 1994).

## RESULTS

We performed a series of statistical analyses to identify the significance of our experimental results. We analysed the experiments separately and compared the abundance of organisms or the weight of periphyton between exclusion zones and control areas.

### *TYPE I EXPERIMENTS*

Type I experiments excluded only shrimps for a period of time and community responses were then observed (Table 2).

Periphyton dry mass decreased over time in the exclusion areas and was significantly less than that in the controls at the end of the experiment in both experiments (Table 2). Algal mass, as measured by chlorophyll a, was not significantly different between treatments in experiment 2, and was almost significantly less in the exclusion of experiment 4, ($p = 0.053$).

Ephemeropterans increased in numbers when shrimps were excluded (Table 2). Electrified treatments had significantly more ephemeropterans (Table 2, $p < 0.05$) than the non electrified treatments for at least the last sampling days of both experiments.

Chironomids appeared to increase in exclusion treatments in the first ten days, but later their density declined. The overall differences between treatments were not significant.

### *TYPE II EXPERIMENTS*

Type II experiments excluded shrimps and inhibited ephemeropterans for a period of time and community responses were then observed (Table 3). Although there were no statistically significant differences between ephemeropteran densities in electrified and non electrified treatments, the former treatment showed an increase in periphyton mass (Table 3). As the main ephemeropteran food resource is periphyton and this resource increased in electrified treatments we concluded that electricity was able to inhibit the foraging of ephemeropterans.

Chlorophyll *a* increased inside type II exclusion zones. However, the increase was statistically significant only for the last sampling day of experiment 1 (Table 3, $p < 0.05$). The other two experiments did not show any significant difference between treatments, but tendencies were the same as in experiment 1 with more chlorophyll *a* inside the exclusion zone. However total dry mass was significantly higher inside exclusion zones for all three experiments for at least the last two sampling days (Table 3, $p < 0.05$).

During experiments 3 and 4 shrimp activity diminished in non-electrified zones and ephemeropterans increased in numbers in those same treatments. There was no significant difference in ephemeropteran numbers between treatments.

Chironomids also responded to Type II experiments. This time the increases were sustained until the end of the experiment. No statistically significant difference was registered between treatments for experiment 1 (Table 3), but experiments 3 and 4 did register a significantly higher density in electrified treatments (Table 3, $p< 0.05$).

### *MODELLING COMMUNITY DYNAMICS*

Interactions among the community components were deduced from the behaviour of the community in our experiments and were reinforced by the literature on their feeding habits. The model describes community interactions under natural and experimental circumstances but does not treat them specifically as trophic relationships. Three models were constructed in order to reproduce community interactions; in natural conditions (Fig.1A), under Type I experimental conditions (Fig.1B) and under Type II experimental conditions (Fig.1C).

When any potential interaction was too small to interfere in the community dynamics, it was not included, as well as any other interaction that could not be observed and measured under experimental conditions. In order to keep the model simple a minimum of parameters was adopted thus reducing the model imprecision.

Total dry mass was included in the model to represent detrital organic matter, actively incorporated organic matter and inorganic accumulation in the periphyton. This includes organic matter of the periphytic algae, but since algae mass is more than $10^3$ times less than total dry mass (Brito *et al*. 2006) it was considered insignificant, permitting the modelling of the two components separately.

We chose to model the system using differential equations because it is a method that is well founded in the literature and is suitable when the variation rate is small. The variation rate of the components of the community was considered small during our experiments, because the daily variation did not surpass ten percent of the total population densities. The system characterizes the animals' community visitation and the algal and detritus accumulation and loss with the equations representing that dynamics. The population is given in densities mean per day.

We used three equation systems to model the community dynamics; the last two models are essentially based in the first. To simulate experimental conditions mathematically we simply excluded equations or the terms that were operating the excluded or inhibited interactions.

Natural Condition

$$\begin{cases} \dfrac{dA_{(t)}}{dt} = r_a A_{(t)} - \left(\dfrac{r_a}{K_a}\right) A_{(t)}^2 - e_a E_{(t)} A_{(t)} \\[2ex] \dfrac{dM_{(t)}}{dt} = r_m - \left(\dfrac{r_m}{K_m}\right) M_{(t)} - e_m E_{(t)} M_{(t)} \\[2ex] \dfrac{dC_{(t)}}{dt} = r_c A_{(t)} - m_c S_{(t)} C_{(t)} \\[2ex] \dfrac{dE_{(t)}}{dt} = r_e A_{(t)} - m_e M_{(t)} E_{(t)} \\[2ex] \dfrac{dS_{(t)}}{dt} = r_s E_{(t)} - s_s S_{(t)}^2 \end{cases}$$

Type I exclusion - shrimps excluded

$$\begin{cases} \dfrac{dA_{(t)}}{dt} = r_a A_{(t)} - \left(\dfrac{r_a}{K_a}\right) A_{(t)}^2 - e_a E_{(t)} A_{(t)} \\[2ex] \dfrac{dM_{(t)}}{dt} = r_m - \left(\dfrac{r_m}{K_m}\right) M_{(t)} - e_m E_{(t)} M_{(t)} \\[2ex] \dfrac{dC_{(t)}}{dt} = r_c A_{(t)} - c_c C_{(t)}^2 \\[2ex] \dfrac{dE_{(t)}}{dt} = r_e A_{(t)} - e_e E_{(t)}^2 \end{cases}$$

Type II exclusion - shrimps excluded and Ephemeroptera nymphs inhibited

$$\begin{cases} \dfrac{dA_{(t)}}{dt} = r_a A_{(t)} - \left(\dfrac{r_a}{K_a}\right) A_{(t)}^2 \\[2ex] \dfrac{dM_{(t)}}{dt} = r_m - \left(\dfrac{r_m}{K_m}\right) M_{(t)} \\[2ex] \dfrac{dC_{(t)}}{dt} = r_c A_{(t)} - c_c C_{(t)}^2 \\[2ex] \dfrac{dE_{(t)}}{dt} = r_e A_{(t)} - e_e E_{(t)}^2 \end{cases}$$

Equations parameters are:

$r_i$ - growth rate of component $i$,

$K_i$ - Logistic term of component $i$,

$e_i$ , $c_i$ , $mi$ and $s_i$ - components interaction coefficients.

*A, M, C, E* and *S* - are the respective densities of algae, Total Dry Mass, chironomids, ephemeropterans and shrimps. Where $dA_{(t)}$, $dM_{(t)}$, $dC_{(t)}$, $dE_{(t)}$ and $dS_{(t)}$ are the mean value of components present in time t + Δt, considering Δt the smallest time interval enough to observe measurable variations.

Type I experiments were simulated by a different equation system, where shrimps were absent and two new terms in the ephemeropteran and chironomid equations substituted the previous terms where shrimps were interacting. This is so because these terms were irrelevant for the community dynamics in the presence of shrimps, but are important now that the major source of population decrease is gone. These terms provide a control for those populations in order not to let them increase in numbers infinitely.

Type II experiments were simulated by a third differential equation system. These conditions prevented ephemeropterans from grazing on algae and total dry mass thus demanding withdrawal of equations terms responsible for this. As only the ephemeropteran foraging activity was inhibited and they were not excluded they still appear in the system.

*Periphyton algae and Total Dry Mass increase rate and Logistic terms*

These constants were calculated via the NONLIN module of SYSTAT statistical computer program using a self consistent test. We used data from Type II exclusion experiments to estimate the increase rates and logistic terms (Table 4).

*Chironomid, ephemeropteran and shrimp increase rate*

Chironomid increase rate was derived from the variation of chironomids in function of algal density (Fig. 2). Ephemeropteran increase rate was calculated from the variation of ephemeropterans in function of algae numbers (Fig. 3) and shrimp increase rate could be due the presence of chironomids or ephemeropterans. We carried out two linear regression analyses plotting the variation of shrimps with the density of Chironomids and the variation of shrimps with the density of ephemeropterans (Fig. 4 A and B). The variation of shrimps was recorded at night and plotted against the densities of the insects registered in the previous afternoon.

*Interaction coefficients of components ($e_i$, $s_i$ and $c_i$)*

These parameters were calculated considering the system equilibrium points. The equation system for natural conditions was considered to be in its stable equilibrium point just before experiments. Thus, substituting the left side of the equations in the natural condition system for zero, we represented the system in its equilibrium point. Then we calculated the interaction coefficients by inserting in the equations the values of the variable before electrification and all the parameters already estimated in the equations.

$$
\begin{cases}
0 = r_a A - \left(\dfrac{r_a}{K_a}\right) A^2 - e_a EA \\
0 = r_m - \left(\dfrac{r_m}{K_m}\right) M - e_m EM \\
0 = r_c A - s_c SC \\
0 = r_e A - s_e ME \\
0 = r_s E - s_s S^2
\end{cases}
$$

Equations parameters are:
$r_i$ - growth rate of component *i*,
$K_i$ - Logistic term of component *i*,
$e_i$ and $s_i$ - components interaction coefficients.
*A, M, C, E* and *S* - are the respective densities of algae, total dry mass, chironomids, ephemeropterans and shrimps at the equilibrium point in natural conditions.

Experiment conditions demanded a new system of equations since shrimps were excluded. The above procedure was repeated to calculate the two interaction coefficients introduced in the ephemeropteran and chironomid equations in order to substitute the shrimp interaction coefficient. In this case we considered the system stable equilibrium point in the last days of Type II exclusion experiments.

$$\begin{cases} 0 = r_a A - \left(\dfrac{r_a}{K_a}\right) A^2 \\ 0 = r_m - \left(\dfrac{r_m}{K_m}\right) M \\ 0 = r_c A - c_c C^2 \\ 0 = r_e A - e_e E^2 \end{cases}$$

Equation parameters are:
$r_i$ - growth rate of component *i*,
$K_i$ - Logistic term of component *i*,
$c_i$ and $e_i$ - components interaction coefficients.
*A*, *M*, *C* and *E* - are the respective numbers of algae, total dry mass, chironomids and ephemeropterans at the equilibrium point in Type II exclusion conditions.

Since there are two experiments for each exclusion type, this allowed us to reach two quantitatively different equilibrium points for each manipulation. Using those values one at a time for parameters calculations led us to two sets of interaction coefficient values for each exclusion type. We calculated the final value for interaction coefficients as the mean of the two sets (Table 4).

*MODEL STABILITY*

The stability of the models was tested by analysis of their eigenvalues. A stable community matrix will return to its previous dynamic equilibrium point after perturbation. However the magnitude of the perturbation must not be too strong since stability in these models means stability in the vicinity of the equilibrium point. This is appropriate for biological purposes since one can understand that it would be impossible for a system to recover to its previous condition in the event of very large perturbations such as the extinction of some of the species. However smaller perturbations are frequent in nature and are mostly not enough to change the system permanently.

For each experimental condition we built its respective community matrix and calculated the eigenvalues. If the eigenvalues turned out to be all negative the model was accepted as a stable model. If at least one of the eigenvalues was positive the model has some instability.

The first community matrix described the community in its natural condition. For this we obtained 5 eigenvalues, all of them negative (Table 5). The second community matrix characterized the community under Type I exclusion and had all 4 eigenvalues negative (Table 5). The last community matrix in which the community was under Type II exclusion conditions also showed negative eigenvalues for all 4 eigenvalues (Table V). All three models were accepted as stable models.

### *MODEL SIMULATION*

We ran the model in the Model Maker 2.0 software program that executed it through a 4-order Runge-Kutta calculation. We made simulations for each model component under all experimental conditions for a time period of 35 days. This time period spans all experiment times including the longest. The model's qualitative and quantitative behaviour was compared to the field data for all model components for both Type I and Type II exclusion conditions.

We used the Type I exclusion equation system to model the community dynamics and compare with the data from the original experiments. Field data of chlorophyll *a* representing algae density was plotted alongside algal density from the simulation. These experiments registered in the exclusion zone a sharp reducing of chlorophyll *a* (Fig. 5A). total dry mass equation model simulation also was plotted alongside field data collected in Type I exclusion experiments (Fig. 5B). The exclusion of shrimps in Type I exclusion experiments caused the ephemeropterans to increase inside electrified treatments. Field data registered for such conditions were plotted with the simulation of Type I conditions and revealed a similar qualitative and quantitative pattern among them (Fig. 6A). On the other hand the chironomid equation model did not perform well when compared to Type I exclusion experiment field data (Fig. 6B). This probably was caused by the high oscillations registered during those experiments for this component in particular. Field data also did not show a distinctive pattern, which did not allow a more accurate model.

Type II experiments excluded shrimps and inhibited ephemeropteran foraging activity, which raised chlorophyll *a*, total dry mass and chironomid density inside the electrified zones. Chlorophyll *a* densities registered during these experiments had a great amount of quantitative variation but the same qualitative response (Fig. 7A). Apart from that, the model simulation fits well quantitatively at least for experiment 3. Total dry mass in Type II exclusion experiments also registered great variability among experiments, although all experiments had the same qualitative response and at least for experiments 3 and 5 the model fits well quantitatively (Fig. 7B).

The quantitative response of chironomids to Type II experiments also varied considerably among experiments, but increasing density inside electrified zones was the overall result. Model simulation approach fitted well for experiment 3 but had less quantitative precision for experiments 1 and 5 (Fig. 8).

### *MODEL PREDICTIONS*

We simulated the extending of the experiment in time to investigate the model's behaviour. The field experiments did not last for more than 20 days, thus any information on how the community dynamics would develop for a longer time period was welcome. The extension in time of Type I exclusion simulation showed that the initial increasing of chironomids and ephemeropterans in the first 20 days was expected to reduce with some oscillation and to stabilize at considerably lower density (Fig. 9). Although the ephemeropteran population stabilized at a higher number than expected in natural conditions, the chironomid population stabilized at much lower values compared to natural densities.

Total dry mass simulation extended in time also showed a surprising pattern. After the initial decrease of the first 20 days, it changed to an increase and then decreased before stabilizing at values lower than those of the natural conditions (Fig. 10). Such oscillation before stability suggests that if nothing more interfered in the community for more than 1000 days of experiment, one should expect total dry mass to increase for nearly 250 days (Fig. 10 for time 150 to 400 days) for no other reason than the community dynamics itself.

Type I exclusion experiments excluded shrimps from the system, which leaves no reason to simulate shrimp population dynamics.

Type II exclusion simulations used the equation system for such circumstances. We also extended these simulations in time, but in this case the system maintains only three components, algae, total dry mass and chironomids. Both algae and total dry mass have fixed logistic terms which cause these populations to stabilize at the established value (Table 4). The time extension modelling allowed us to predict chironomid population size when stabilized (Fig. 11). Chironomid, ephemeropteran and shrimp populations do not have fixed logistic terms, which allows their dynamics to stabilize in different population numbers if the algae logistic term is changed.

## DISCUSSION

### *EXPERIMENT RESULTS*

Type I exclusion experiment successfully excluded shrimps from the electrified zone (Tab. II). As a consequence, ephemeropteran densities increased in the electrified zones for both experiment 2 and 4. Under natural conditions we observed that ephemeropterans in our study site are diurnal and shrimp activity is mostly nocturnal (Moulton *et al.* 2004). Such antagonism suggested that ephemeropterans were avoiding shrimps in time, probably to avoid predation (Moulton *et al.* 2004). Ephemeropterans have been shown to avoid trout predators by diurnal movements (Peckarsky & McIntosh 1998). Experiment results are in accordance with the predator avoidance hypothesis since in the absence of shrimps, ephemeropterans tended to increase and even prolong their activity in electrified zone into the night (Moulton *et al*. 2004).

Ephemeropteran foraging activity is mainly directed at the periphyton growing on the bedrock. Ephemeropterans are known as important grazers in some systems (Hill & Knight 1987). We observed decreases in chlorophyll *a* and total dry mass of the periphyton in both Type I experiments in which ephemeropterans increased, and we deduce that ephemeropterans are important grazers in our system. Chironomid densities oscillated greatly during Type I experiments. The exclusion of shrimps was not enough to distinguish the effects of ephemeropterans from the effect of chironomids. The logical extension of the experiments was to exclude ephemeropterans and observe the effects of chironomids in Type II experiments

We managed to inhibit ephemeropterans foraging activity inside the electrified zones, but were not able to prevent them from entering the exclusion zones in Type II experiments (Table 3). When we observed ephemeropterans directly inside high-intensity (Type II) electrical exclusion areas, we saw that they were affected by the electrical pulses – they twitched and would often jump or lose their hold on the substrate. Direct visual counts showed that their numbers diminished in the exclusion areas (Silveira 2002, Moulton *et al*. 2004). Thus we were surprised that the counts of ephemeropterans sampled by Surber apparatus did not reflect the direct observations. This was at least partially due to the presence of small ephemeropterans, which were probably less affected by the shocks and were less visible. However we concluded that the high-intensity electricity inhibited their foraging, since periphyton increased in the exclusion areas. Brown *et al*. (2000) reported a similar phenomenon with high intensity exclusion – macroinvertebrate numbers did not decrease in the excluded areas, but grazing pressure was apparently diminished.

We were able to separate the ephemeropteran effect from the chironomid effect, since ephemeropteran foraging activity was inhibited, and that of the chironomids apparently not. Chironomid effect on periphyton turned out to be negligible since chlorophyll *a* and total dry mass increased greatly despite the increase of chironomids (Fig. 7). It was not possible to extract the effects of chironomids since we had no means to exclude the very small chironomids with electricity.

Type II exclusion registered significant increase for both total dry mass and chironomid densities, as well as the significant exclusion of shrimps. Chlorophyll *a* results were significant only for experiment 1, but showed the same increasing tendency for experiments 3 and 5. The explanation for the lack of significance of chlorophyll *a* is not clear but might be associated with the low concentration of chlorophyll *a*, low sensitivity of the spectrophotometric method of measurement and possible variation in the primary productivity due to seasonal effects.

The qualitative results of our experiments were quite consistent with a trophic cascade interaction system (Moulton *et al*. 2004). Trophic cascades are expected to occur where primary producers are exposed to herbivore pressure of a few grazer species which in turn have few but strong predators (Strong 1992). Moreover, low spatial heterogeneity encourages trophic cascades. Our community has these characteristics and vindicates Strong's predictions. The phenomenon we found is an interaction cascade and not entirely trophic, since the effects of shrimps over ephemeropterans was more frequently of behaviour nature than due to ingestion itself. The result is quite different to those found in electrical exclusion experiments with similar shrimp and insect fauna in Puerto Rico (Pringle & Blake 1994) and Costa Rica (Pringe & Hamazaki 1998), where the exclusion of shrimp produced increases in periphyton. Pringle & Hamazaki (1998) suggested that the lack of a trophic cascade was due to the omnivorous trophic status of the shrimps. Our result implies that the shrimps were primarily predators of ephemeropterans and chironomids.

Although the experiments were not internally replicated, we repeated each manipulation at least twice, which gives us confidence in the experimental results. Based on this confidence, we used the data for our modelling exercise.

*MODELLING THE COMMUNITY INTERACTIONS*

The experimental data set provided a useful base for modelling the community since it was able to show the most significant qualitative relationships, gave us estimates of quantitative responses and fulfilled mathematical demands for calculating model parameters.

The experimental design, however, was not the most appropriate for mathematical modelling. The time lag between the first and the next sampling days was larger than desirable for calculating accurate growth rates. However, the experimental area had little room to accommodate replicated treatments and not enough sampling space inside the treatments to admit more than three or four sampling days.

Despite these shortcomings, the manipulation experiments proved essential for developing a dynamic view of the community interactions. This allowed us to build models that involved all the identifiable interactions, one for each community condition. All three models retained the same parameter mainframe in order to achieve a more universal description of the community interactions.

A difference equation system model might have been used since our samplings days are discrete in time and variations in experiment effects among these samples are not discernable. However, we decided to employ differential equation systems because they are less dependent on the given initial condition. Modelling our system with differential equations could have reduced the quantitative precision of the simulations. However, the initial conditions of the community components showed a considerable amount of variation in our experiments, which was considered to be more pertinent than the possible loss of quantitative precision.

All three models proved to be stable by eigenvalue analysis. Each matrix, for each modelled condition had negative eigenvalues. Community matrices with negative eigenvalues are characteristic of systems that are stable in the vicinity of their equilibrium points. This means that the system is able to recover from small perturbations and return to its previous dynamic equilibrium. This is expected for most natural systems, due to their natural ability to recover from small perturbations (Pimm 1991). It can be noted, however, that stream ecosystems are subject to frequent perturbations due to floods. This has given rise to non-equilibrium models of stream communities (e.g. McAuliffe, 1984) and investigations of patchiness as determinants of community composition (e.g. Downes *et al*. 1993).

Classic Lotka-Volterra equations of population dynamics consider that population variation is due to a balance between increase (birth, migration and death rate included) and decrease, which is generally, described as intra-specific competition (Begon *et al*. 1986). Berryman *et al*. (1995) went further by adding some equation terms describing trophic relationships. However these systems are suited for communities which the major force for population increase is reproduction. A previous attempt at modelling the community dynamics based on Berryman-like models showed the need for a more specific set of models (Silveira & Moulton 2000), since the main force for population increase in our experiments was migration and not reproduction.

In the case of our community, only the algae population increased due to reproduction, while chironomid, ephemeropteran and shrimp population growth rates were due to migration. Although algae population growth rate could be set as proportional to algae population, the other three groups had their population increase rate proportional to their potential prey. This simulates immigration due to an attraction force. Moreover, as the model describes the most important interactions, it is not imperative that a relationship appears in both directions. For instance ephemeropterans have effect on total dry mass described in the equation systems of natural condition and Type I exclusion but do not have any effect from total dry mass in their equation. This is so because the grazing activity of the ephemeropterans affects periphyton total dry mass but their real interest lies in the algae stock mixed with the detritus matrix (Krsulovic 2004); stable isotope analysis of ephemeropterans showed them to be herbivores of algae despite their large effect on the non-algal parts of the periphyton (Brito *et al*. 2006).

The dynamics of total dry mass was a special case with a steady accumulation rate not proportional to anything. Its logistic term represents the periphyton loss due to hydrological sheer stress. Another interesting feature of our models is that chironomids, ephemeropterans and shrimps do not have a fixed logistic term. This is more likely since they are dependent on the amount of algae or ephemeropterans and not dependent on an arbitrary maximum density. Algae population and total dry mass density are, however, most likely dependent on nutrients and hydrology respectively, and these could be understood as naturally determined limits.

The Type I exclusion simulation is in good agreement with both qualitative and quantitative experimental results. Simulation of chlorophyll *a*, total dry mass and ephemeropterans fitted well with experiment data (Fig. 5 and 6). However the high quantitative variability of the experimental results is probably the main reason for the poor simulation of chironomids in Type I exclusion conditions. Such high quantitative variability could be explained by a transient gradient expected to be observed under these circumstances. Hastings (2001) emphasizes that when nature is submitted to new conditions a transient dynamics is expected before a new community quantitative and qualitative composition stabilizes. Short term experiments are criticized because the observed results could change in a longer period of time. Modelling with field data can help simulating long term experiments and comparing the simulation with the short term experiment results. We did find differences between our observed results and the long term experiment simulated by our models, which implies a possible effect of short term experiments.

The simulation of Type II exclusion also approximated the dynamics of chlorophyll *a*, total dry mass and chironomids (Fig. 7 and 8). However these experiments registered an even larger amount of quantitative variability with experiments 1 and 3 showing high discrepancy between their chlorophyll *a* – total dry mass relationship. Such difference among experiments suggests that there are some relevant processes being overlooked by our experimental approach.

One of the most striking results of the modelling appeared when we simulated the extension in time of our experiments. Type I exclusion model predicts that the initial increase of chironomids and ephemeropterans stabilizes in much lower numbers compared to the amount of these insects in 30 days of experiment. Algae density never reaches zero and stabilizes at low numbers. Ephemeropterans stabilize at higher numbers than natural concentrations but chironomids are reduced in density compared to the natural condition (Fig. 9).

It is particularly interesting that the simulation of Type I exclusion predicts that chironomids should decrease. They are sessile animals and depend on periphyton as habitat as well as food. The model did not anticipate the decrease as an intrinsic property of the chironomid equation, since the equation is essentially the same as that of the highly mobile ephemeropterans, which increased. The model construction based on field data highlighted chironomid dependence on periphyton substrate (Souza *et al.* 2007).

Simulation allowed us to see further in time and, according to the model, it shows that results observed during the experiments are expected to change. Type II exclusion modelling extended in time was useful in suggesting where chironomid population density would stabilize, a prediction that could not be achieved solely by experiment observations (Fig. 11).

Time extension modelling also suggests that Total dry mass density under Type II exclusion is expected to change if the experiment could last longer. An increase is expected when the experiment is about 150 days of duration and then it turns to decrease again around 400 days of experiment (Fig. 10). The oscillation happens in consequence of the reduction of ephemeropterans which also reduce their grazing pressure on periphyton, thus allowing total dry mass to increase for some time. If such time extension could be observed in the field and nothing more than the community dynamics interfered in the system such increasing would imply a possible important element that had been overlooked.

## *CONCLUSIONS*

Manipulation experiments are very appropriate for modelling purposes. Some considerations must be taken first relative to experimental design and sampling. Experimental design and sampling must fit the modelling needs to provide more adequate results for modelling. Nevertheless, even the less adequate experimental designs can always produce a qualitative model.

Our models did represent the system well in both qualitative and quantitative aspects. The experimental design showed some shortcomings that, allied with high natural variability, did take away some of the model's quantitative precision.

The modelling exercise helped to determine the most important interactions and their probable nature and discard the less important community interactions previously established in the first model for that community (Silveira & Moulton 2000). Such result offers a more clear and objective understanding of the community dynamics.

Interesting results came up, suggesting that densities of ephemeropterans and chironomids in Type I exclusion were expected to be lower, that chironomids density could be higher in Type II exclusions and Total dry mass could oscillate before stabilizing in Type II exclusions.

Modelling with field data predicts the biological need of chironomids for substrate fixation in Type I exclusion. The simulation of Type I exclusion extended in time shows their high dependence on substrate.

The models can be useful for any other circumstances where the population balance is not determined by birth and death but is determined by migration. Such conditions are common for field experiments as well as management programs.

**Table 1.** Experiment characteristics. Shrimp exclusion = Type I exclusion and Shrimp exclusion and Ephemeropterans inhibition = Type II exclusion.

| Experiment | Date (month/year) | Exclusion type | Fauna affected | Experiment duration in days | Sampling days | Treatment dimensions |
|---|---|---|---|---|---|---|
| 1 | May,1998 | II | Shrimp and Ephemeropterans | 13 | 0, 7, 9, 13 | 60 X 200 cm |
| 2 | July, 1999 | I | Shrimp | 19 | 0, 5, 12, 19 | 60 X 200 cm |
| 3 | October, 1999 | II | Shrimp and Ephemeropterans | 34 | 0, 6, 13, 34 | 60 X 200 cm |
| 4 | January, 2000 | I | Shrimp | 15 | 0, 6, 15 | 30 X 180 cm |
| 5 | January, 2000 | II | Shrimp and Ephemeropterans | 15 | 0, 6, 15 | 30 X 180 cm |

**Table 2.** Type I exclusion results. ↑ = increased density, ↓ decreased density and * = p<0.05

| Exclusion type | Total Dry mass | Chlorophyll *a* | Chironomid | Ephemeropteran | Shrimps |
|---|---|---|---|---|---|
| Exp. N° 2 Type I exclusion | ↓* | ↓ | Did not change | ↑* | ↓* |
| Exp. N° 4 Type I exclusion | ↓* | ↓ | Did not change | ↑* | ↓* |

**Table 3**. Type II exclusion results: ↑ = increased density, ↓ decreased density and * = p<0.05

| Exclusion type | Total Dry mass | Chlorophyll *a* | Chironomid | Ephemeropteran | Shrimps |
|---|---|---|---|---|---|
| Exp. N° 1 Type II exclusion | ↑* | ↑* | Did not change | Did not change | ↓* |
| Exp. N° 3 Type II exclusion | ↑* | ↑ | ↑* | ↑ | ↓* |
| Exp. N° 5 Type II exclusion | ↑* | ↓ | ↑* | ↑ | ↓* |

**Table 4**. Model parameters. Units for the ephemeropterans, chironomids and shrimps are based on density of individuals. Algae and total dry mass units are in µg and g.

| Relationship | Coefficient | Parameter |
|---|---|---|
| Total Dry Mass increase rate / day | $r_m$ | 1.95 |
| Algae increase rate/ day | $r_a$ | 0.29 |
| Ephemeropteran increase rate/ day | $r_e$ | 0.122 |
| Chironomid increase rate/ day | $r_c$ | 0.087 |
| Shrimp increase rate/ day | $r_s$ | 0.042 |
| Ephemeropteran effect on Total Dry Mass | $e_m$ | 0.00046 |
| Ephemeropteran effect on algae | $e_a$ | 0.00029 |
| Shrimp effect on ephemeropterans | $s_e$ | 0.032 |
| Shrimp self-interaction | $s_s$ | 0.066 |
| Shrimp effect on chironomids | $s_c$ | 0.0031 |
| Ephemeropteran self-interaction | $e_e$ | 0.000002 |
| Chironomid self-interaction | $c_c$ | 0.000002 |
| Total dry mass Logistic term | $K_m$ | 33.5 g |
| Algae Logistic term | $K_a$ | 1679 µg |

**Table 5**. Eigenvalues analysis results

| Community condition | Eigenvalues |
|---|---|
| Natural | -0.2004<br>-0.0434<br>-0.2854<br>-0.7582<br>-1.4441 |
| Type I exclusion | -0.9506<br>-0.0063<br>-0.2895<br>-0.0174 |
| Type II exclusion | -0.0582<br>-0.0100<br>-0.1260<br>-0.2900 |

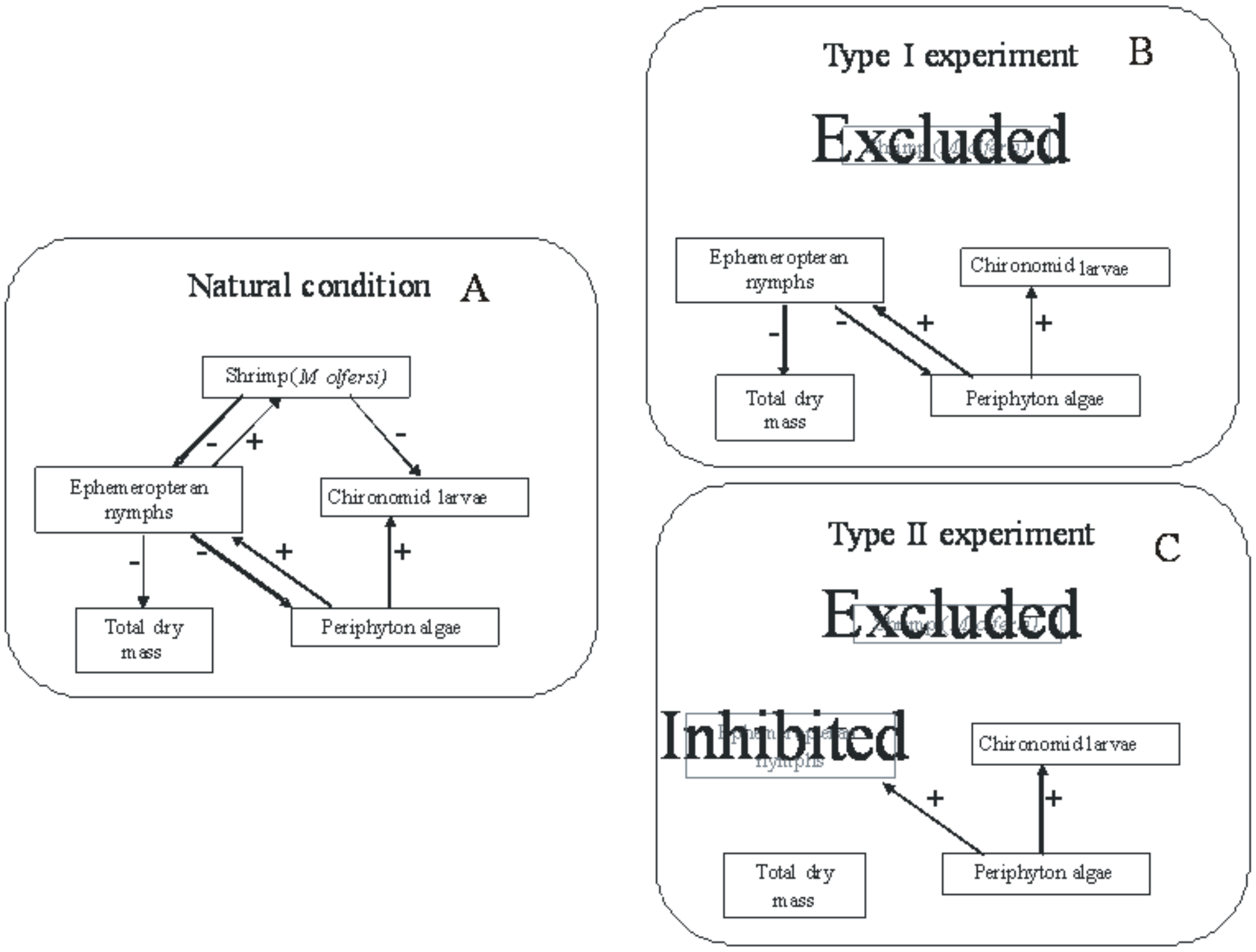

Fig.1. Conceptual model of the community interaction for natural circumstances (A), Type I experiment conditions (B) and Type II experiment conditions (C). Arrows indicate interactions and thus can appear pointing both directions meaning positive (+) and negative (-) interactions. Total dry mass comprises organic and inorganic matter of the periphyton.

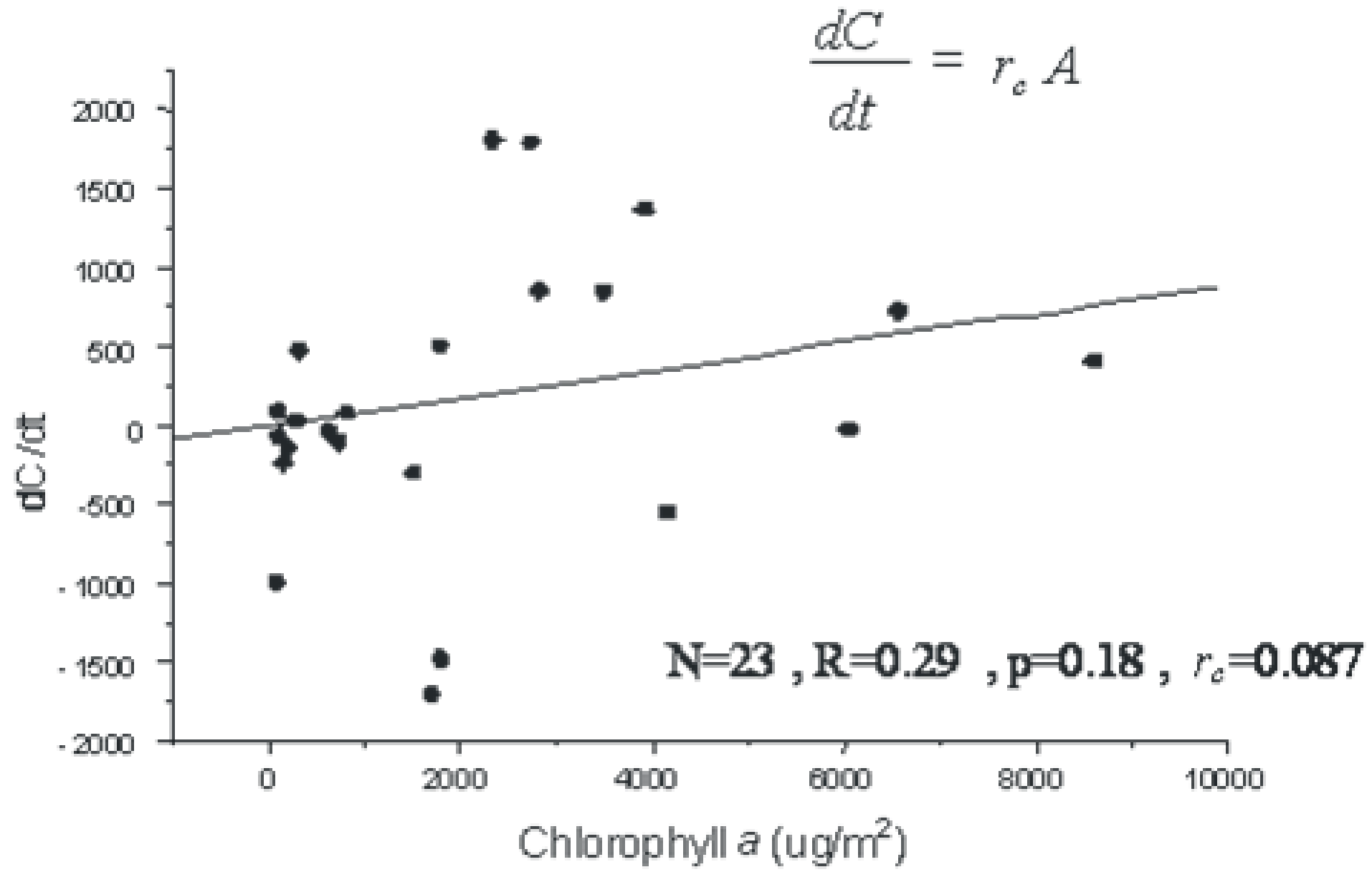

Fig. 2- Linear regression between chironomid variation and algae density for Type I and II exclusion experiments first sampling days.

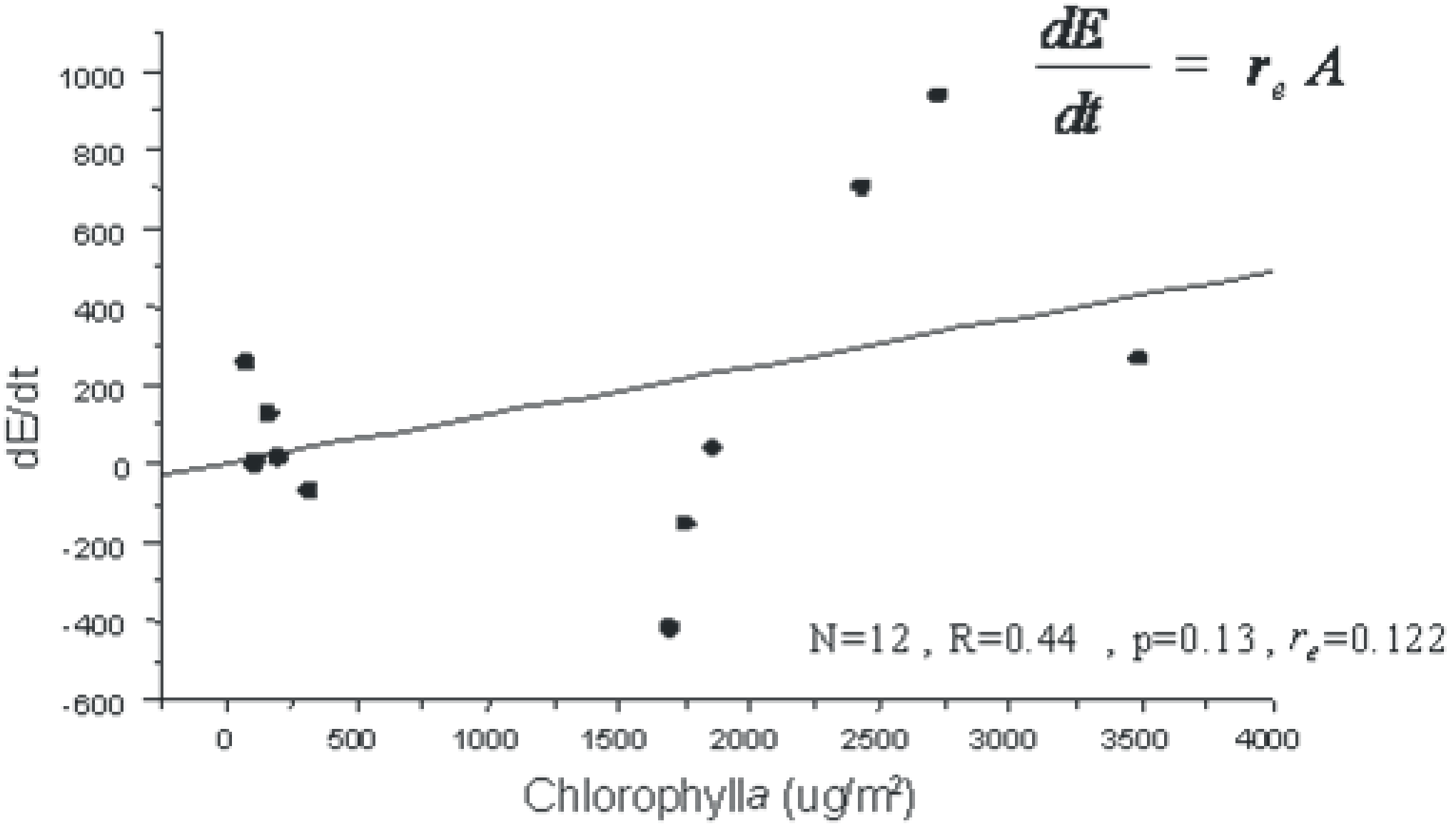

Fig. 3- Linear regression between ephemeropteran variation and algae density for Type I exclusion experiments first sampling days.

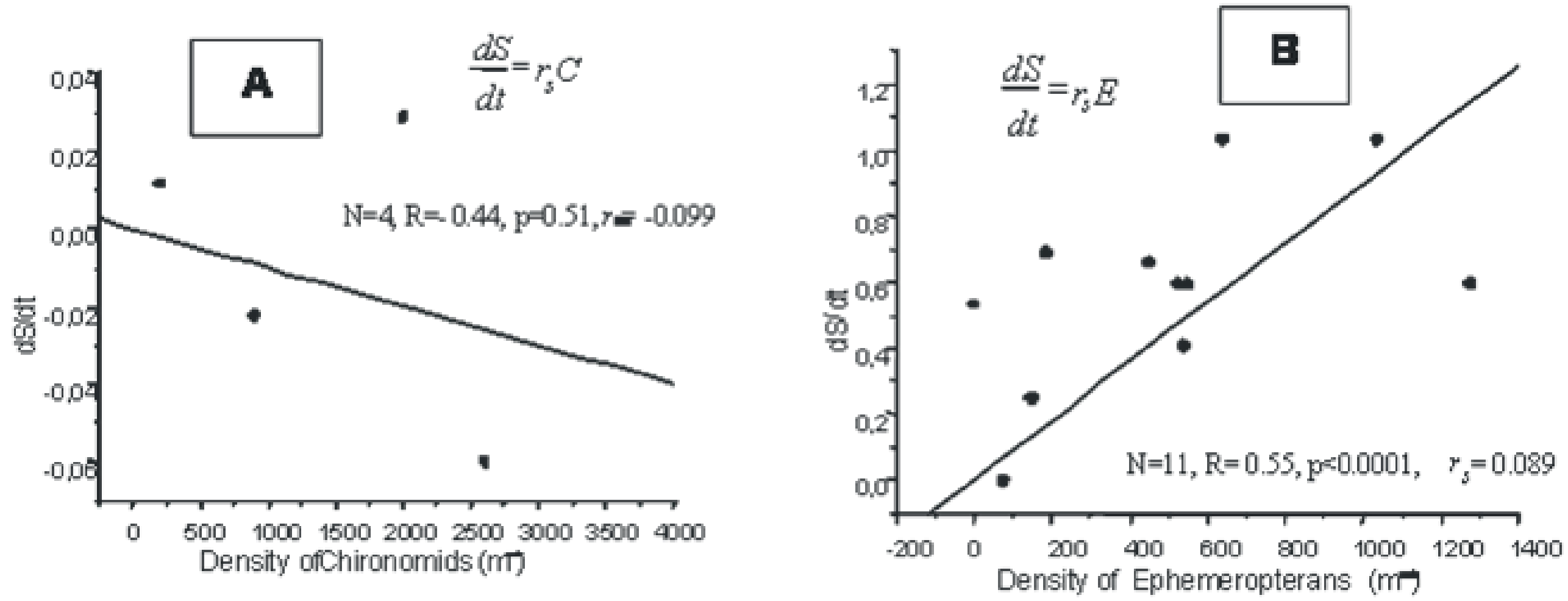

Fig. 4- Linear regression between shrimp variation and chironomid density (A) and ephemeropteran density (B) for Type I and II exclusion experiments first sampling days.

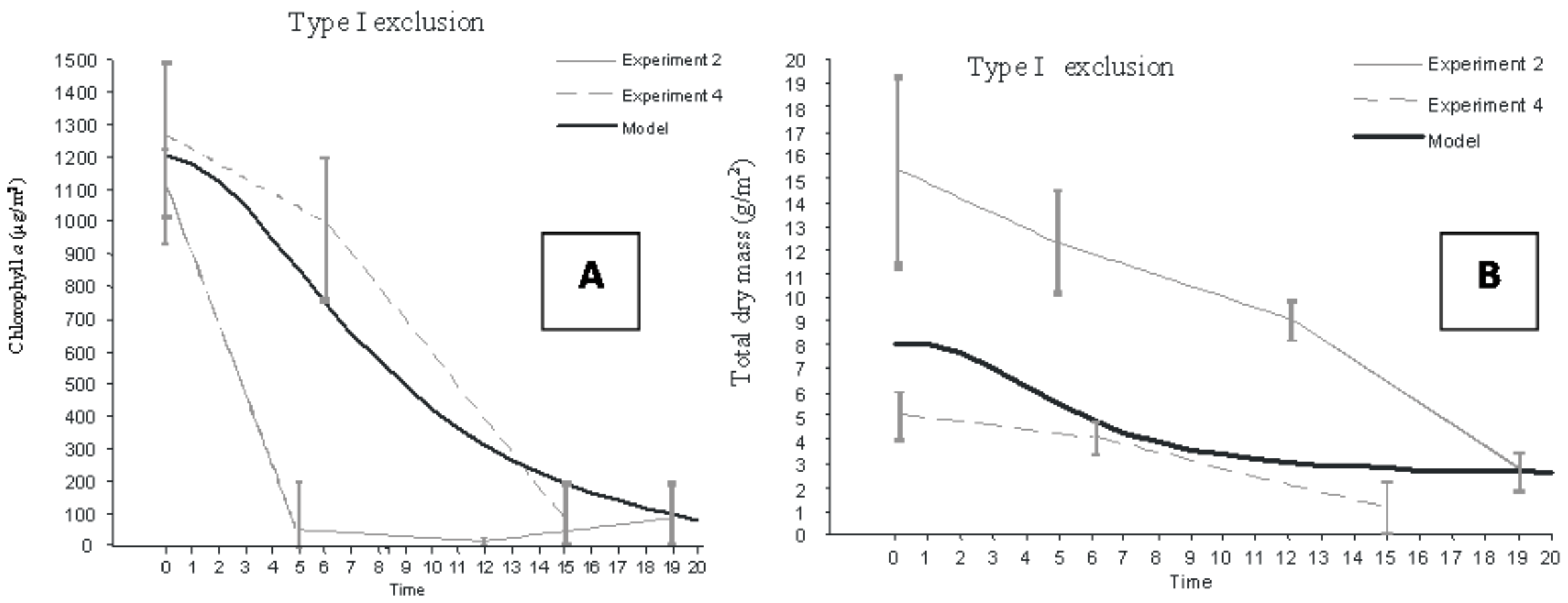

Fig. 5- Comparison of field experimental data and model simulation for Type I exclusion. Chlorophyll *a* (A) and total dry mass (B).

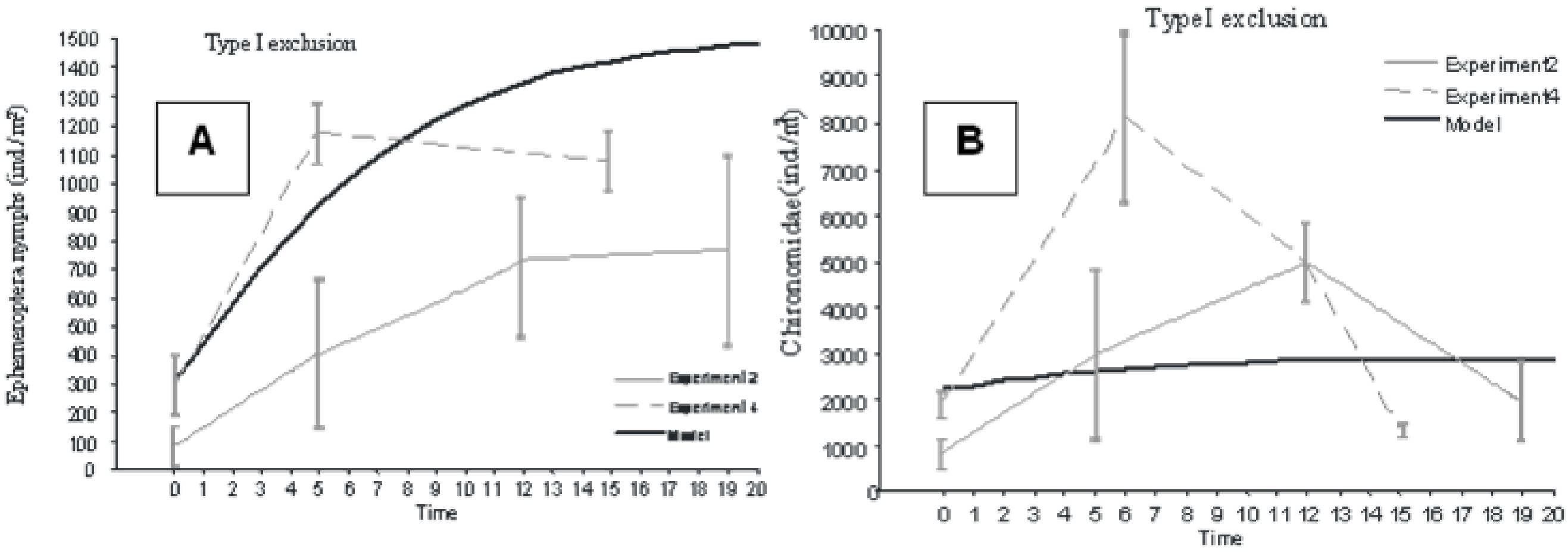

Fig. 6 - Comparative results of field experiment data and model simulation. Ephemeropteran (A) and chironomid (B) dynamics during Type I exclusion experiments.

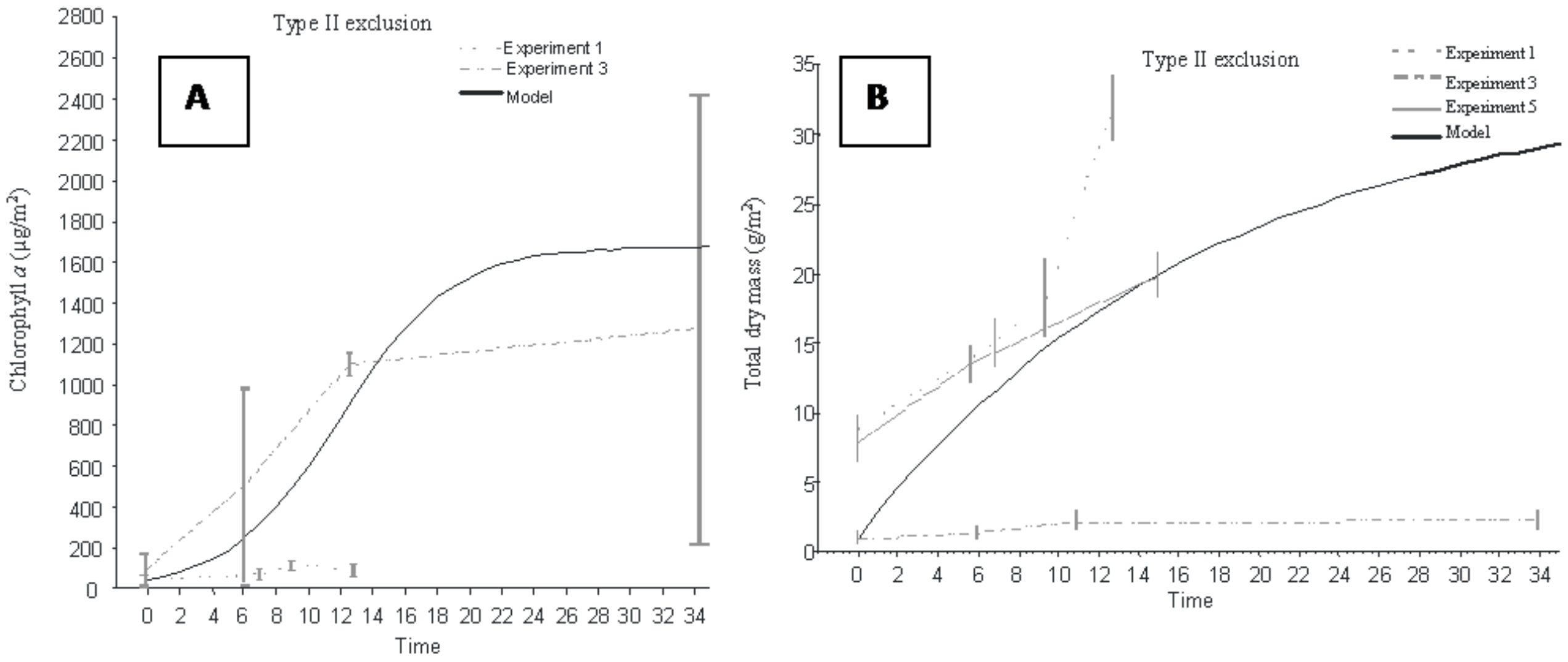

Fig. 7 - Comparative results of field experiments data and model simulation. Chlorophyll *a* increasing during Type II exclusion experiment (A) and Total dry mass response during Type II exclusion experiments (B).

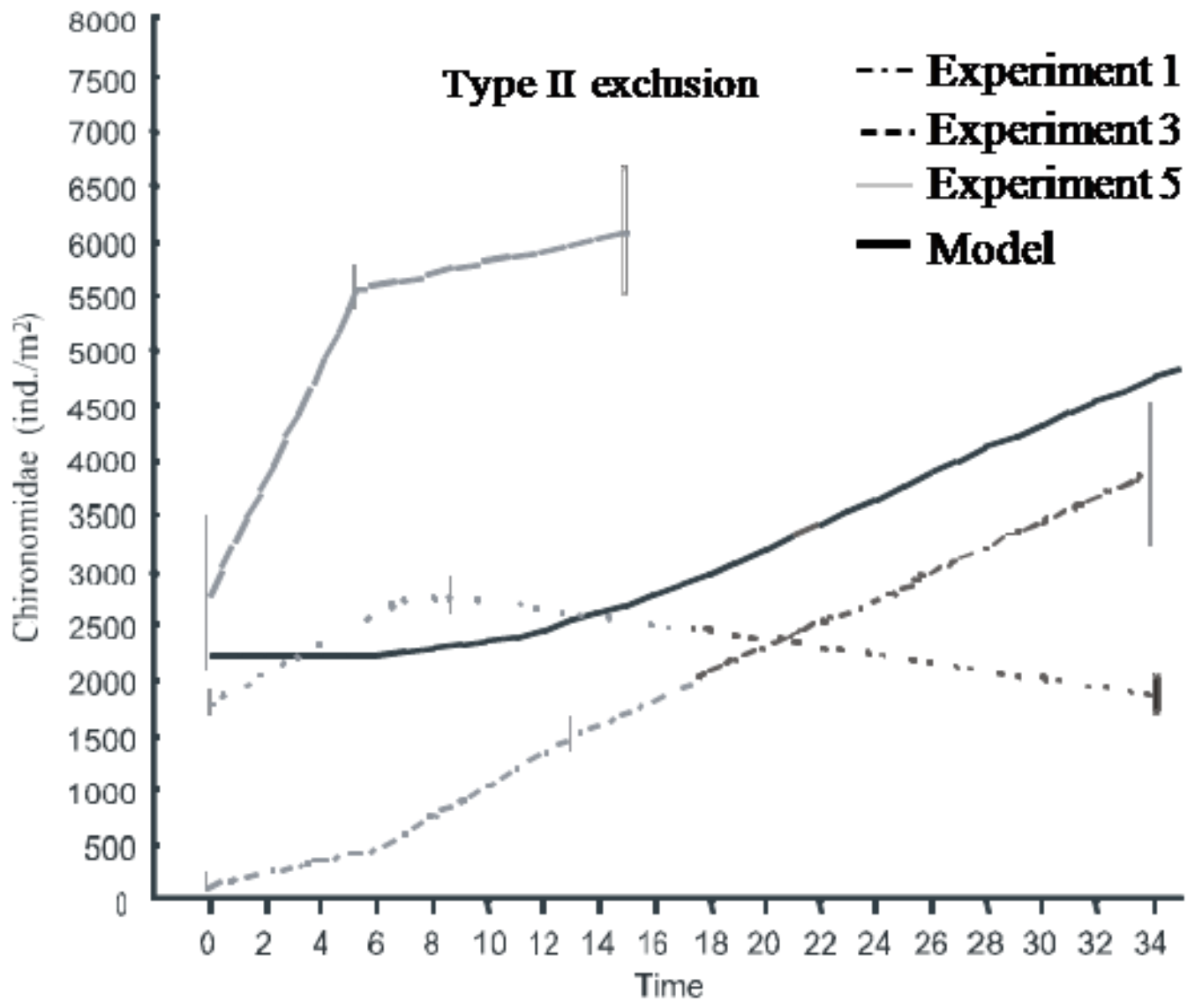

Fig. 8 - Comparative results of field experiment data and model simulation. Chironomid response during Type II exclusion experiment.

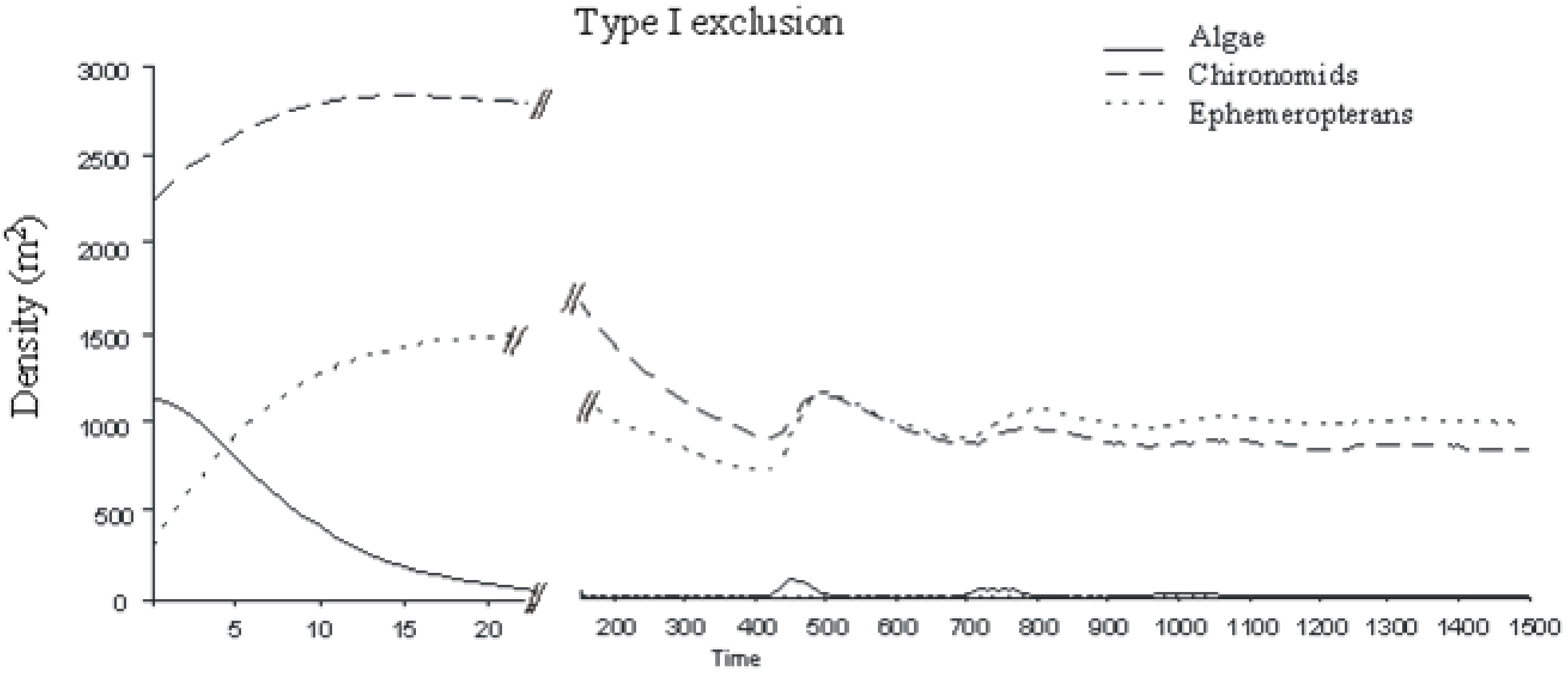

Fig. 9 – Simulation of Type I exclusion modelling extended in time. The initial values for all components are considered those for community stable equilibrium in natural conditions.

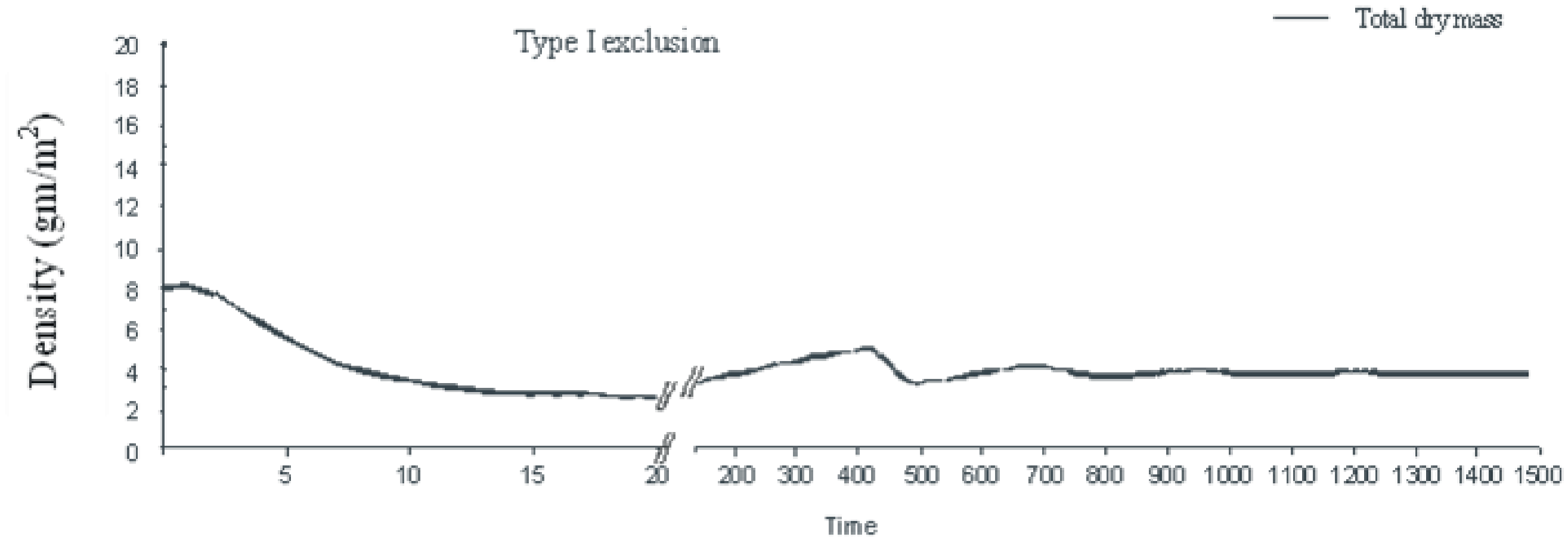

Fig. 10 – Simulation of Type I exclusion modelling extended in time. The initial value for total dry mass is considered as the community stable equilibrium in natural conditions.

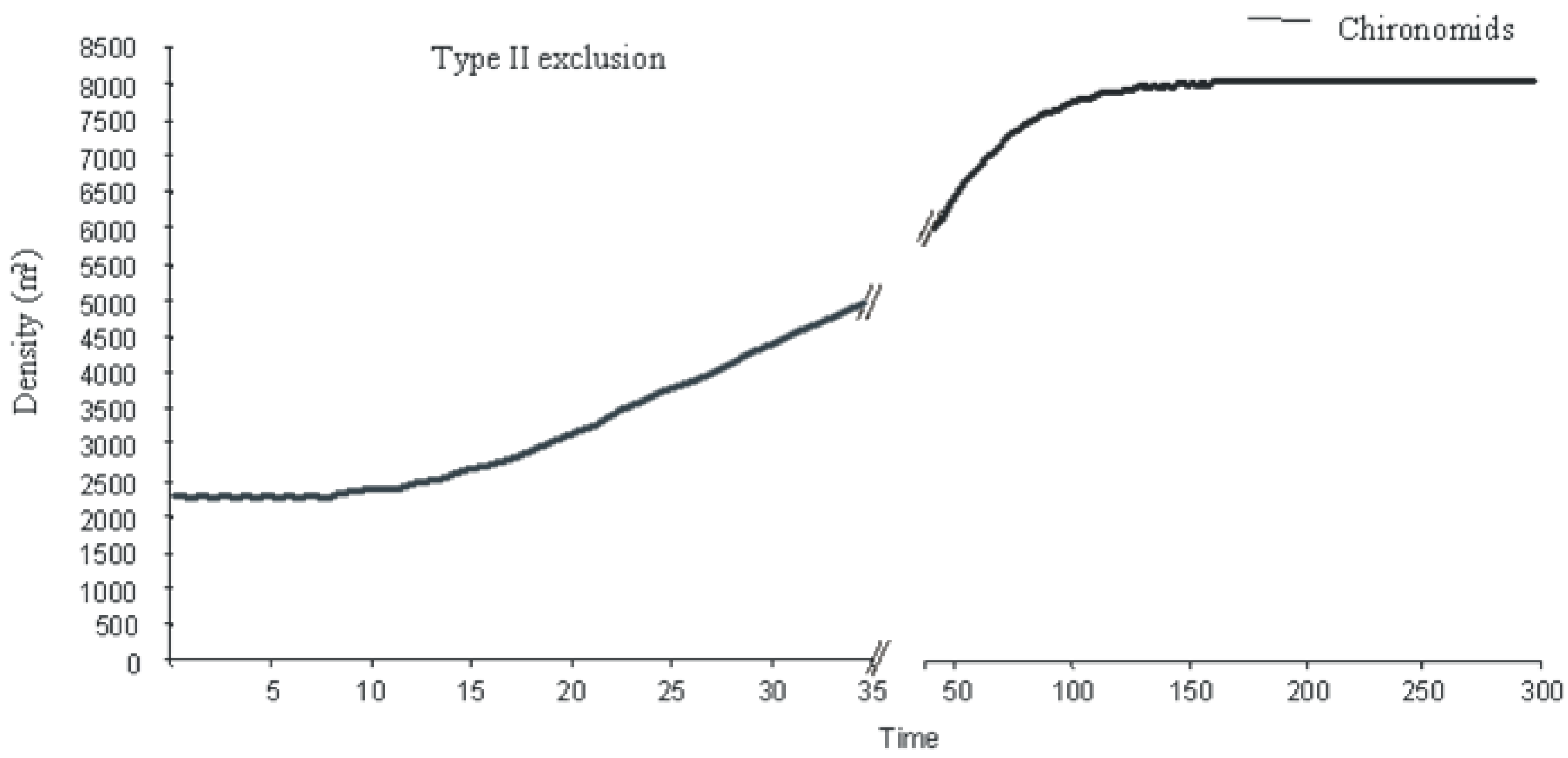

Fig. 11 – Simulation of Type II exclusion modelling extended in time. The initial value for chironomids is the community stable equilibrium in natural conditions.